\documentclass[12pt]{iopart}  \usepackage{graphicx}
\usepackage{subfigure,cite} \usepackage{amsfonts,amssymb}

\newcommand{\text}[1]{\mbox{\scriptsize{#1}}}

\begin{document}

\title[Non-Equilibrium Dynamics of Single Polymer
Adsorption]{Non-Equilibrium Dynamics of Single Polymer Adsorption to
Solid Surfaces}

\author{Debabrata Panja$^*$, Gerard T. Barkema$^{\dagger,\ddagger}$
and Anatoly B. Kolomeisky$^{**}$}

\address{$^*$Institute for Theoretical Physics, Universiteit van
Amsterdam, Valckenierstraat 65,\\ 1018 XE Amsterdam, The Netherlands

$\dagger$ Institute for Theoretical Physics, Universiteit Utrecht,
Leuvenlaan 4,\\  3584 CE Utrecht, The Netherlands

$^{\ddagger}$Instituut-Lorentz, Universiteit Leiden, Niels Bohrweg 2,
2333 CA Leiden,\\ The Netherlands

$^{**}$Department of Chemistry, Rice University, Houston, TX 77005 USA}

\begin{abstract} Adsorption of polymers to surfaces is crucial for
understanding many fundamental processes in nature. Recent
experimental studies indicate that the adsorption dynamics is
dominated by non-equilibrium effects. We investigate the adsorption of
a single polymer of length $N$ to a planar solid surface in the
absence of hydrodynamic interactions. We find that for weak adsorption
energies the adsorption time scales $ \sim N^{(1+2\nu)/(1+\nu)}$,
where $\nu$ is the Flory exponent for the polymer. We argue that in
this regime the single chain adsorption is closely related to a
field-driven polymer translocation through narrow pores. Surprisingly,
for high adsorption energies the adsorption time becomes longer, as it
scales $\sim N^{(1+\nu)}$, which is explained by strong stretching of
the unadsorbed part of the polymer close to the adsorbing surface.
These two dynamic regimes are separated by an energy scale that is
characterised by non-equilibrium contributions during the adsorption
process.
\end{abstract}


\maketitle

Polymer adsorption is a fundamental phenomenon that controls many
natural processes \cite{fleer_book}. The adsorption of the polymeric
molecules to different surfaces and interfaces is important for
adhesion, colloidal stabilisation, development of composite materials
and coatings, for cell adhesion and communication, and for protein-DNA
interactions \cite{fleer_book,lodish_book}. The importance of polymer
adsorption has motivated extensive experimental and theoretical
investigations to understand the underlying mechanisms. As a result,
the equilibrium properties of adsorbed polymers are now
well-understood \cite{degennes87}. However, many experimental studies
\cite{johnson92}, supported by theoretical ones
\cite{shaffer93,raviv02,review05} indicate that non-equilibrium
behaviour is increasingly important in polymer adsorption dynamics.

One key parameter in polymeric adsorption is the height of the free
energy barrier that monomers have to overcome in order to bind to the
surface. If the barrier is high, one commonly calls the adsorption
process chemisorption, while in the absence of a significant barrier,
it is called physisorption. A further characterization of physisorption
involves the strength of the binding interaction between each
monomer and the surface. If this interaction is on the order of $kT$,
the process is called weak physisorption, while one speaks of strong
physisorption in the case of interactions of about 10 $kT$ or more, as
for instance typically encountered for hydrogen bonding.

In chemisorption, the high barrier faced by monomers attaching
to the surface slows down the adsorption process; this allows the
adsorbed part of the polymer chain to partially relax in effectively
equilibrium conformations, giving rise to formation of large loops via
the accelerated zipping mechanism \cite{review05,SV03}.  The absence
of a significant barrier makes non-equilibrium effects even more
important in physisorption \cite{johnson92,review05}. It is not clear
what mechanisms drive the polymer adsorption away from equilibrium in
this regime \cite{review05}.  One of the possible contributions is the
interaction between neighbouring polymer molecules that can significantly
slow down the overall dynamics.  This source of deviation from equilibrium
is commonly eliminated by considering the adsorption dynamics of single
polymers \cite{review05,shaffer94,ponomarev00,oshaughnessy1,vaneijk98}.

The adsorption of single macromolecules for weak polymer-surface 
interactions has been investigated by a combination of analytical and 
computational methods 
\cite{review05,shaffer94,ponomarev00,oshaughnessy1,vaneijk98,descas06,bhatta08}. 
Monte Carlo simulations with the bond fluctuation model revealed 
significant deviations from equilibrium dynamics 
\cite{shaffer94,ponomarev00}.  The adsorption time was reported to scale 
as $\sim N^{1.57 \pm 0.07}$ for self-avoiding polymers, while the 
exponent is equal to $1.50 \pm 0.04$ when the excluded volume 
interactions are neglected. Computer simulations and an approximate 
theory were also used to investigate irreversible adsorption of tethered 
chains \cite{descas06,bhatta08}. These investigations assumed that the 
polymer molecule during the adsorption has three parts: a segment of 
already bound monomers, a stretched linear part (``stem'') and a 
remaining part (``flower'') which is not affected by the force of 
adsorption. This theoretical model yields an adsorption time scaling as 
$\sim N^{\alpha}$ with $\alpha=1+\nu \approx 1.59$. Here $\nu$ is the 
Flory exponent for the polymer, and $\nu\approx0.588$ in three 
dimensions. Simultaneously, in the Monte Carlo simulations a smaller 
value of $\alpha$, namely $\approx 1.51$, has been observed 
\cite{bhatta08}, but it was argued that finite-size effects were 
responsible for this discrepancy. The stem-flower model was originally 
proposed by Brochard-Wyart \cite{brochard95} for polymer chains under 
strong flows (under constant and very large flow velocity). It has a 
clear physical picture that allows one to obtain specific predictions 
for the dynamical properties. However, the growth velocity of the 
adsorbed polymer has been shown to be not large 
\cite{descas06,bhatta08}, not constant, and in time it even decays to 
zero; hence the validity of the stem-flower model to adsorption in all 
situations is questionable.  Thus, despite many attempts, mechanisms of
the single-polymer binding to the surfaces are still not well-understood.
In this paper we present theoretical arguments supported by simulation 
data that clarify several non-equilibrium features of single-polymer 
adsorption.
\begin{figure}[h]
\begin{center}
\includegraphics[width=0.5\linewidth]{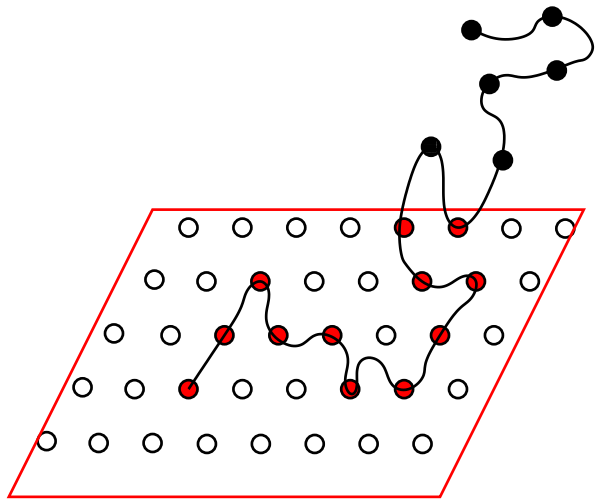}
\end{center}
\caption{Schematic view of the single polymer adsorption to a planar
surface. Unfilled circles correspond to binding sites available for
adsorption by monomers. Red filled circles describe the surface sites
already occupied by the polymer. Black filled circles represent
monomers that are not adsorbed on the surface yet. \label{fig1}}
\end{figure}

The polymer molecule during binding to the surface can be viewed as
consisting of two segments: the adsorbed monomers and the block of
free monomers not on the surface. Theoretical studies argue that
adsorption (for weak interactions) can be viewed as a sequential
zipping process \cite{descas06,bhatta08} in which the size of the
adsorbed block increases by one monomer a time. This sequential
mechanism suggests that the single-polymer adsorption process is
closely related to field-driven polymer translocation (detailed later
in the text), which has been intensively studied in recent years
\cite{luo08,vocks08}. During field-driven translocation, the polymer
molecule moves through a pore sequentially, decreasing the number of
monomers on the {\it cis\/} side of the pore and sequentially
increasing the number of monomers on the {\it trans\/} side of the
pore. Theoretical studies of translocation based on the microscopic
dynamics of the polymer \cite{vocks08,trans} showed that memory
effects are crucial for understanding this process. The memory effects
appear due to the finite time to dissipate away or replenish the local
enhancement in the density of monomers at the pore. From these works
on translocation, it is reasonable to expect that the same memory
effects in the polymer should also play a role in the adsorption of
single polymers to a surface.

Consider a single polymer molecule that is near a solid surface, which
starts to adsorb to the surface as shown in Fig. \ref{fig1}. We assume
that there are uniformly distributed binding sites on the surface,
that the energy of adsorption per site is equal to $\varepsilon$, and
that the distance between binding sites is the same as the size of
each monomer. We then use a Monte Carlo based FCC-lattice polymer code
in three dimensions for self-avoiding polymers, with the rigid flat
surface placed at $z=0$, and study the adsorption dynamics for a
variety of polymer lengths and for different strengths of the
adsorption energy, in the absence of hydrodynamic interactions. In
this polymer model the individual monomers perform both reptation and
``sideways'' movements \cite{model} with each kind of movement
attempted with frequency unity, which provides us with the definition
of time. This model has been used before to simulate the diffusion and
exchange of polymers in an equilibrated layer of adsorbed
polymers~\cite{klein_adsorbed}. Recently, we have used this code
extensively to study polymer translocation under a variety of
circumstances \cite{wolterink06,vocks08,trans}.

The process of adsorption involves a change in the free energy $\Delta
F$ for the polymer: the polymer loses energy due to the attractive
interaction between the surface and the monomers, and loses entropy,
as adsorption makes the polymer collapse into two dimensions from
three. The value of the adsorption energy $\varepsilon$ dictates the
sign of $\Delta F$, and thereby determines the fate of a partially
adsorbed polymer. At high adsorption energies, the polymer will nearly
completely adsorb since adsorption is energetically favoured, while at
low adsorption energies, it will desorb since desorption is
entropically favoured. In between, there is a critical value
$\varepsilon^*$ at which a partially adsorbed polymer will both adsorb
and desorb with equal probability, for which the entropy gain for
desorption is exactly compensated by the energetic gain for adsorption.

Consider a partially adsorbed polymer of length $N$, with $m$ monomers
(counting from one end) completely adsorbed on the surface while the
rest $(N-m)$ monomers are moving freely (off the surface).  If one
assumes that the adsorbed part of the polymer takes the form of a
self-avoiding random walk on the two-dimensional adsorbing plane, then
the partition function of this polymer is given by $Z=[A_2\mu_2^m
m^{\gamma^{\text(2D)}-1}][A_3\mu_3^{(N-m)}
(N-m)^{\gamma_1^{\text(3D)}-1}]$. Here, $\gamma^{\text(2D)}=49/32$ and
$\gamma_1^{\text(3D)}\approx1.16$ are two universal scaling exponents
\cite{diehla}, and $A_2$, $A_3$, $\mu_2$ and $\mu_3$ are
model-dependent quantities. For this partially adsorbed polymer
$\displaystyle{\frac{\partial\Delta F(\varepsilon)}{\partial m}}$ can
be estimated as $\displaystyle{\frac{\partial\Delta
F(\varepsilon)}{\partial m}}\simeq
-\varepsilon+k_BT\ln(\mu_3/\mu_2)+1/N\,\,\mbox{corrections}$.
Equating $\displaystyle{\frac{\partial\Delta F(\varepsilon)}{\partial
m}}$ to zero then yields the critical adsorption energy
$\varepsilon^*\simeq k_BT\ln(\mu_3/\mu_2)$ in the limit of
$N\rightarrow\infty$.  Note that this expression is only an estimate,
since in the adsorbed state not all monomers of the polymer adhere to
the surface; moreover, as has been demonstrated in
Ref. \cite{descas06}, the adsorbed part of the polymer takes a very
compact conformation --- much more compact than a self-avoiding walk
in two dimensions. Nevertheless, $\displaystyle{\frac{\partial\Delta
F(\varepsilon^*)}{\partial m}}=0$ shows that $\varepsilon^*$ is in the
order of $k_BT$.

For our model we determine $\varepsilon^*$ in the following manner. We
start with a polymer of length $N$ with $N/2$ monomers from one end
constrained to the surface (i.e., constrained to $z=1$) without an
adsorption energy, while the remaining $N/2$ monomers are free, and
equilibrate the polymer under this constraint (the free $N/2$ monomers
encounter the surface only as a planar obstacle). At time $t=0$, an
adsorption energy $\varepsilon$ is introduced, and simultaneously the
constraint is lifted. We repeat this exercise for polymer lengths
ranging from $N=100$ to $N=800$, while tuning the suppression of the
desorption rate by a factor of $v\equiv \exp[-\varepsilon/(k_BT)]$,
until on average the adsorbed part of the polymer neither grows nor
shrinks.  The results for the critical values $v^*$ for several
polymer lengths are summarised in Table \ref{table1}; from this Table
we conclude that $v^*\approx 0.34\pm 0.01$ for our model, and thus
that $\varepsilon^*/(k_BT)=1.08 \pm 0.03$. Since we use
$\varepsilon\ge2k_BT$, our polymers always adsorb, and any reference
to high or low adsorption energies will henceforth refer to
$\varepsilon>\varepsilon^*$.
\begin{table}[h]
\begin{center}
\begin{tabular}{cc} $N$&$v^*$\tabularnewline \hline\hline
100&0.405\tabularnewline 120&0.395\tabularnewline
140&0.392\tabularnewline 160&0.38\tabularnewline
200&0.365\tabularnewline 400&0.36\tabularnewline
800&0.342\tabularnewline \hline\hline
\end{tabular}
\caption{The critical desorption rate
$v^*=\exp[-\varepsilon^*/(k_BT)]$ as a function of polymer length
$N$.\label{table1}}
\end{center}
\end{table}

The specific manner in which we simulate surface adsorption is as
follows. We take a polymer of length $(N+n^*)$ with $n^*$ monomers
from one end constrained to $z=1$ without an adsorption energy (a
process we term ``grafting'' for later reference), and equilibrate the
rest of the polymer in $z>0$, i.e., during the equilibration process
the $N$ free monomers encounter the surface only as a planar
obstacle. We index the monomers consecutively along the chain,
starting with $i=-n^*$ for the grafted end. The free end is thus
indexed by $i=N$, and the last grafted monomer corresponds to
$i=0$. At $t=0$ we switch on the attractive interaction between the
monomers and the surface, and simultaneously lift the constraint. The
dynamics of the polymer for $t>0$ is then governed by, in addition to
self-avoiding polymer dynamics, the fact that the ratio of probability
of a monomer (including the grafted monomers) jumping from $z=1$ to
$z=2$ and that of a monomer jumping from $z=2$ to $z=1$ is given by
the Boltzmann ratio $\exp[-\varepsilon/(k_BT)]$. Throughout this paper
we choose $n^*=30$; since we use adsorption energies higher than $2
k_BT$, this implies that the probability for the entire polymer to
detach from the surface is practically zero. It should be noted also
that the specific value of $n^{*}$ does not affect the adsorption
dynamics as long as $n^{*} \ll N$.

Given this setup, {\it on average\/} we expect the monomers to be
adsorbed on the surface in a sequential zipping manner: {\it on
average monomer $n_1$ $(>0)$ will be adsorbed on the surface $($i.e.,
attain $z=1)$ for the first time earlier than monomer $n_2>n_1$}. For
future reference, at any time $t$ for any configuration of the
adsorbing polymer we can identify the monomer with the highest index
$n(t)$, which has $z=1$, to be called the ``active monomer''. This
definition divides the entire polymer into two segments: (i) a part
consisting of monomers $i\leq n(t)$, largely adsorbed to the surface,
and (ii) another part consisting of monomers $i>n(t)$ that behave as a
polymer of length $[N-n(t)]$ tethered on the surface at the location
of the active monomer.

In light of $n(t)$ defined in the above paragraph, it is important to
note that our setup involving the initial grafting of the polymer ---
albeit simplified --- does capture the adsorption dynamics in a real
situation. In reality, a long polymer does not adsorb starting from one end;
almost always it starts to adsorb somewhere in the middle. Imagine,
for a polymer of length $N$, a situation when the monomer with index
$n_0$ (somewhere in the middle of the polymer), is the first one to
adsorb. By definition, this monomer immediately divides the polymer
into two separate ``sub-polymers'' --- of lengths $n_0$ and $(N-n_0)$
respectively. If $\varepsilon>\varepsilon^*$, then (on average) these
two sub-polymers will start being adsorbed independently (in stating
this, we disregard the steric interactions between them) {\it from
their common end}: one from monomer $n_0$ towards monomer $1$, and the
other from monomer $n_0$ towards $N$. The adsorption dynamics for a
polymer in a real situation --- at least in the scaling sense, which
is the main focus of this paper --- is the same as that of the polymer
in our setup (which starts to adsorb from one end). Our setup ---
similar to the existing ones \cite{descas06,bhatta08} --- therefore,
is purely a choice of convenience to study the adsorption dynamics in
a real situation.

Returning to our setup, if adsorption were a sequential zipping
process for every single realisation, then the adsorption dynamics can
be described solely by the active monomer index $n(t)$ as a function
of time, and the dynamics of adsorption can be mapped exactly on to
that of field-driven translocation. More precisely, in polymer
translocation driven by a potential difference $\Delta V$ across the
pore, when a monomer crosses from the {\it cis\/} ({\it trans}) to the
{\it trans} ({\it cis\/}) side, the length of the polymer segment on
the {\it cis\/} side reduces (increases) by one monomer with an energy
gain (penalty) of magnitude $q \Delta V$, where $q$ is the charge of
one monomer. Similarly, (a) if the active monomer happens to detach
from the surface (with an energetic penalty $\varepsilon$) then the
length of part (ii) of the polymer increases by roughly one monomer;
(b) alternatively, if the index of the active monomer increases by one
(with an energy gain of $\varepsilon$), then the length of part (ii)
of the polymer decreases by one monomer. In Ref. \cite{vocks08}, based
on memory effects in polymer dynamics, two of us showed that the total
number of translocated monomers at time $t$ increases as a power-law
$\sim t^{(1+\nu)/(1+2\nu)}$ at {\it weak\/} fields; recently, this has
been confirmed by a different polymer models \cite{luo08,fyta}. This
implies that if the adsorption process were a sequential zipping
process for every single realisation for our setup, $n(t)$ would scale
$\sim t^{(1+\nu)/(1+2\nu)}$. Based on this result --- although in a
real situation adsorption is a sequential zipping {\it only on
average, and not for every single realisation} --- on average we
expect $n(t)$ to increase in time $t$ also as $\sim
t^{(1+\nu)/(1+2\nu)}$ for our setup, i.e., the adsorption time scales
as $\sim N^{(1+2\nu)/(1+\nu)}$, when the adsorption energies are not
very high. We demonstrate this in the paragraphs below.

It is important to note that during adsorption for our setup, a
monomer with index $[n(t)+n_1]$ may get adsorbed with the surface
before any of the in between monomers [with indices
$n(t)+1,\ldots,n(t)+n_1-1$] do. For such an event, the adsorbed part
of the polymer is said to form a ``loop'' of length $n_1$ between
monomers with indices $n(t)$ and $[n(t)+n_1]$; in fact, it is
precisely such ``loop formations'' that prevent adsorption --- unlike
translocation, for which the first passage of the monomers through the
pore takes place strictly sequentially --- from being a sequential
zipping process for every single realisation. Consequently, the
traditional way to follow the progress of adsorption for our setup is
to track the average total number of adsorbed monomers $s(t)$ at time
$t$, so that $s(\tau_{ad})\sim N$ would define the adsorption time
$\tau_{ad}$. However, since $s(t)$ for any single realisation will
saturate at a value $\sim O(N)$, care needs to be taken in measuring
$s(t)$, otherwise saturation effects might affect the numerical
determination of the true exponent. In order to avoid saturation
effects, we define $t_n$ as the average time, and $s_n$, as the
average number of adsorbed monomers when the $n$-th monomer attains
$z=1$ for the first time, with the condition that no monomer with
index $>n$ has ever attained $z=1$. Since $t_n$ is defined only till
$n=N$, this method ensures that $s_n$ never saturates.

\vspace{5mm}
\begin{figure}[!h]
\begin{center}
\includegraphics[angle=270,width=0.6\linewidth]{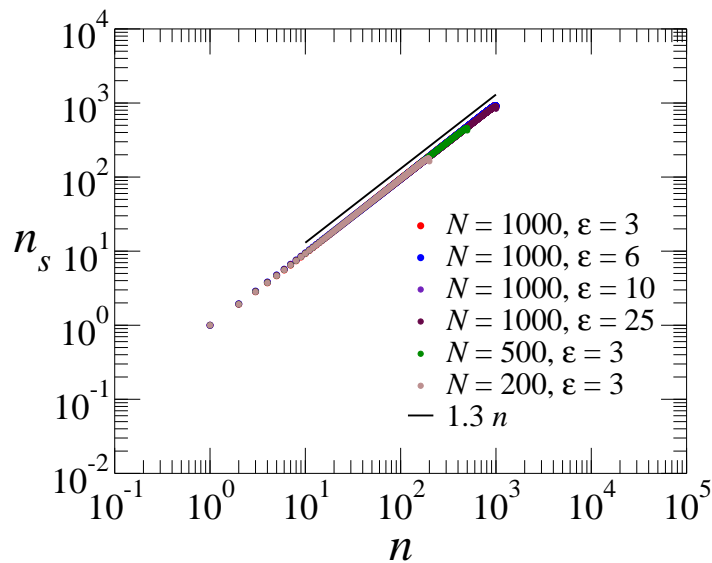}
\end{center}
\caption{Scaling of $s_n$ as a function of $n$, for several values of
$N$ and $\varepsilon$; all curves collapse on a single master curve,
corresponding to the scaling $s_n\sim n$, represented by the solid
black line.\label{fignsvsn}}
\end{figure}

Indeed, we find that adsorption of the individual monomers is {\it
not\/} a sequential process for every single realisation precisely
because of the loop formations as discussed in the above paragraph;
however, as shown in Fig. \ref{fignsvsn}, $s_n$ does scale linearly
with $n$, confirming that {\it adsorption on average is indeed a
sequential zipping process}. This property ensures that the exponent
we get for $s_n$ as a function of $t_n$ is the same as the one that
one would get from tracking $s(t)$ as a function of $t$. A remarkable
feature of Fig. \ref{fignsvsn} is the collapse of all $s_n$ vs. $n$
curves on a single master curve: it shows that the proportion of
monomers in the loops within the adsorbed part of the polymer, given
by $(n-s_n)$, is independent of $\varepsilon$, a feature that we will
return to shortly.

\begin{figure*}
\begin{minipage}{0.33\linewidth}
\begin{center}
\includegraphics[width=\linewidth]{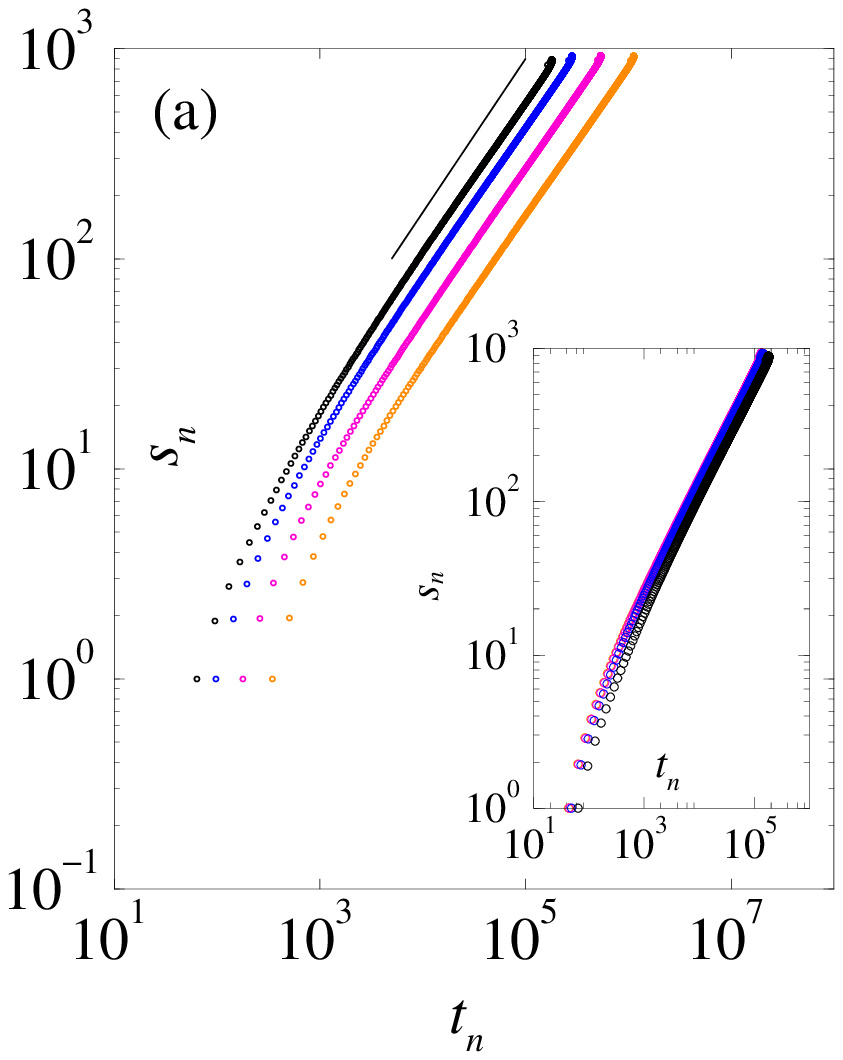}
\end{center}
\end{minipage}
\begin{minipage}{0.33\linewidth}
\begin{center}
\includegraphics[width=\linewidth]{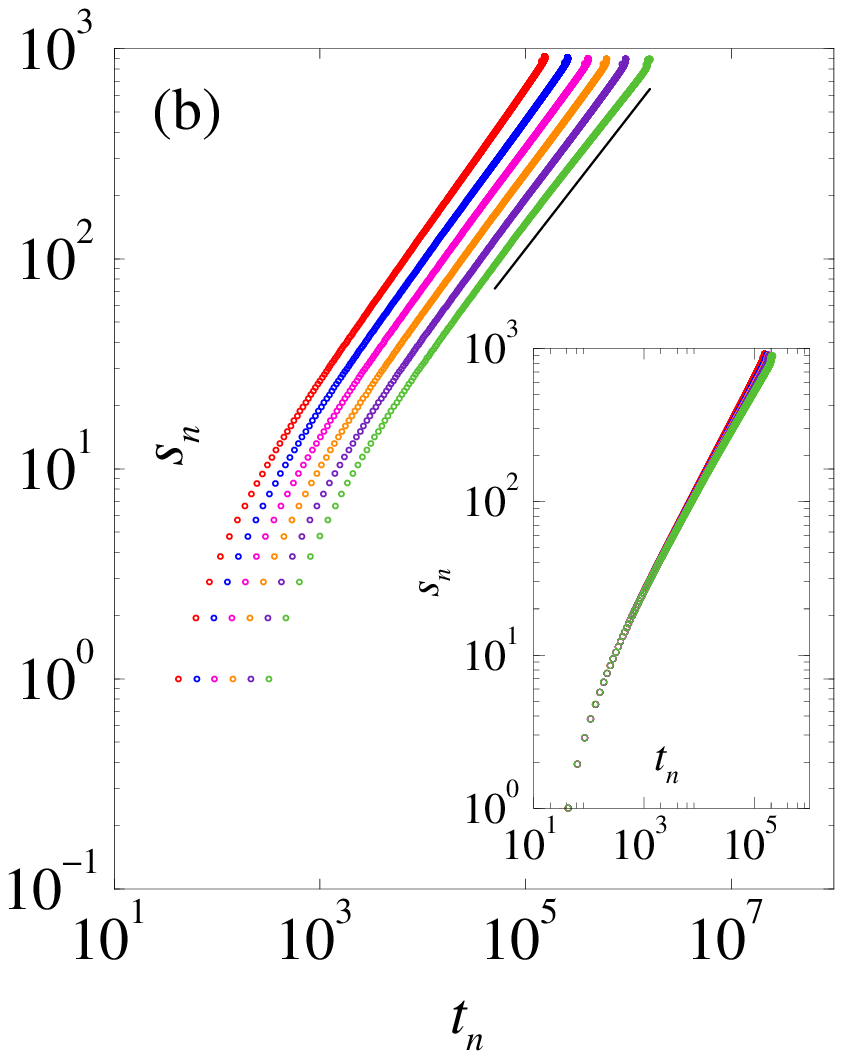}
\end{center}
\end{minipage} \hspace{-4mm}\begin{minipage}{0.34\linewidth}
\begin{center}
\includegraphics[angle=270,width=0.8\linewidth]{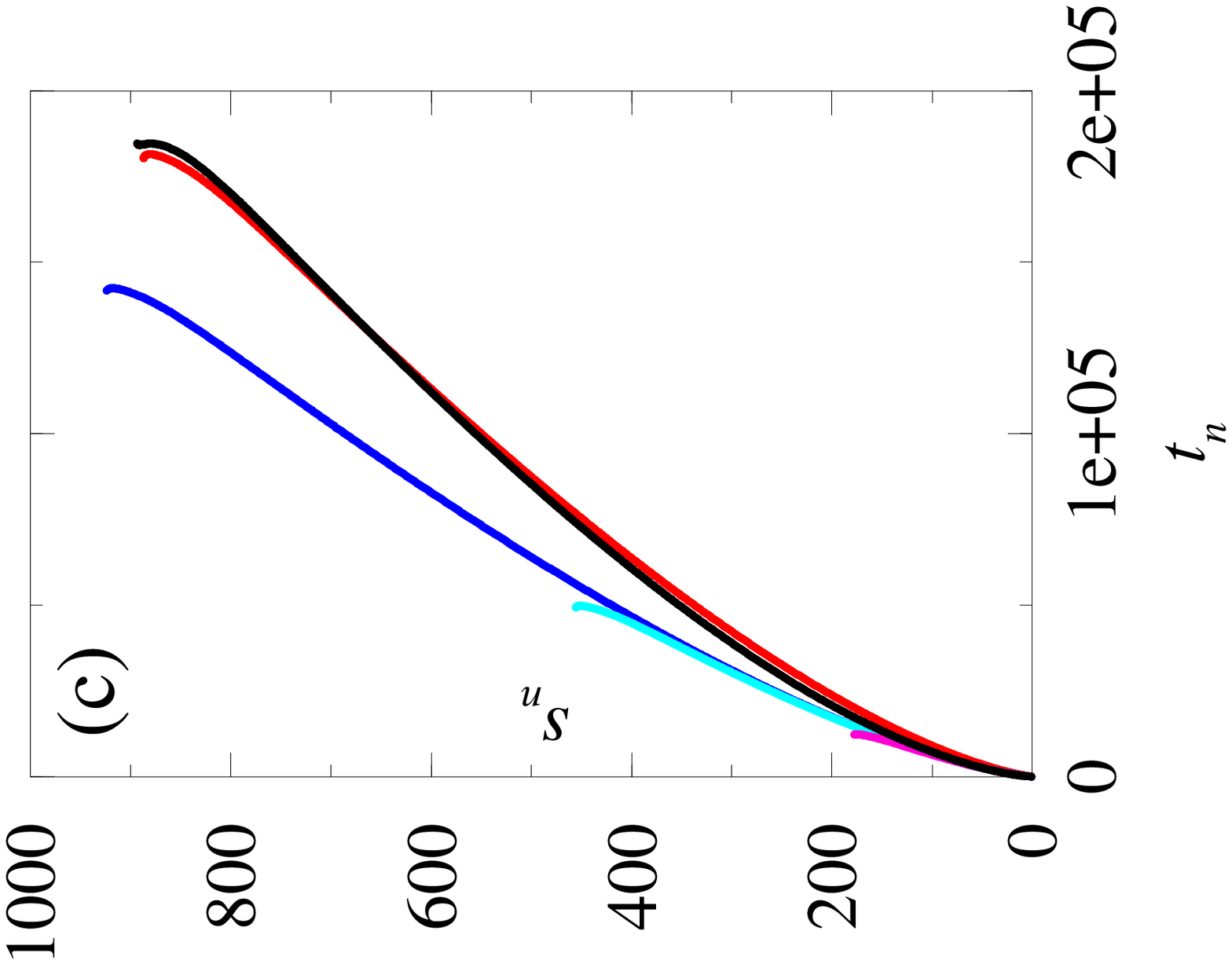}
\end{center}
\end{minipage}
\caption{(a) Weak adsorption data for $N=1000$ (from left to right)
$\varepsilon=2$ (black), $\varepsilon=3$ (blue), $\varepsilon=4$
(magenta) and $\varepsilon=5$ (orange): the data are progressively
separated by a factor $2$ along the $x$-axis for clarity. The original
data are shown in the inset in the same colour scheme. The solid black
line corresponds to an exponent $(1+\nu)/(1+2\nu)\simeq0.73$. (b)
Strong adsorption data for $N=1000$ (from left to right)
$\varepsilon=6$ (red), $\varepsilon=7$ (blue), $\varepsilon=8$
(magenta), $\varepsilon=9$ (orange), $\varepsilon=10$ (brown) and
$\varepsilon=25$ (green): the data are progressively separated by a
factor $1.5$ along the $x$-axis for clarity. The original data are
shown in the inset in the same colour scheme. The solid black line
corresponds to an exponent $1/(1+\nu)\simeq0.63$. Data correspond to
an average over $\simeq400,000$ realisations for each value of $N$ and
$\varepsilon$. Right panel: Collapse of the data for $s(t_n)$
vs. $t_n$ for $\varepsilon=3$, and $N=200$ (red), $N=500$ (blue) and
$N=1000$ (magenta). (c) Comparison of adsorption speed at several
energies and $N$-values: $N=1000$, $\varepsilon=2$ (red), $N=1000$,
$\varepsilon=3$ (blue), $N=500$, $\varepsilon=3$ (cyan), $N=200$,
$\varepsilon=3$ (magenta), and $N=1000$, $\varepsilon=10$
(black). Note that adsorption is slower  for large $\varepsilon$ than
for small $\varepsilon$.\label{fig2}}
\end{figure*}

In Fig. \ref{fig2}(a) we present the data for weak interactions with
the surface, i.e., when the adsorption energy is not too high
($\varepsilon\le5$), for which we do obtain the exponent
$(1+\nu)/(1+2\nu)$, corresponding to $\tau_{ad}\sim
N^{(1+2\nu)/(1+\nu)}$. Additionally, the data exhibit
energy-dependence [see the inset, and also Fig. \ref{fig2}(c)],
demonstrating that for $\varepsilon\le5$ the higher adsorption energy
also yields faster adsorption, like higher field means shorter
(field-driven) translocation time at weak fields \cite{vocks08}. The
situation changes for stronger interactions ($\varepsilon>5$): in
Fig. \ref{fig2}(b), we register a slowly decreasing slope in the
$t_n$-$s_n$ log-log plot with increasing adhesion energy; eventually
for the virtually irreversible adhesion process $\varepsilon=25$, we
recover an exponent $1/(1+\nu)$, i.e., $\tau_{ad}\sim N^{1+\nu}$, in
agreement with
Refs. \cite{shaffer94,ponomarev00,descas06,bhatta08}. For these values
of $\varepsilon$, $\tau_{ad}$ is independent of $\varepsilon$.

The surprising aspect of two different scaling regimes for $s_n$
vs. $t_n$, as shown in Figs. \ref{fig2}(a) and (b) is that for long
enough polymers adsorption is faster for low $\varepsilon$-values than
for high $\varepsilon$-values. This is demonstrated in
Fig. \ref{fig2}(c). In fact, Fig. \ref{fig2}(c) leaves one wondering
whether the slowdown of adsorption is due to phenomena at the
adsorbing surface that are different for strong adsorption energies
than for weak adsorption energies. E.g., for high $\varepsilon$-values
it has been shown in Ref. \cite{descas06} that the adsorbed part of
the polymer takes a very compact form. Based on this result of
Ref. \cite{descas06}, it may be argued that since for high
$\varepsilon$-values the individual monomers are essentially
irreversibly adsorbed, the polymer needs to form systematically bigger
loops to access available surface sites for adsorption, a
phenomenology that is absent for low $\varepsilon$-values; and as a
result adsorption is slower for high $\varepsilon$-values than for low
$\varepsilon$-values. Such a possibility is however ruled out by the
collapse of the data over a very wide range of $\varepsilon$ values in
Fig. \ref{fignsvsn}: it shows that on average the fraction of monomers
in the loops [given by $(n-s_n)$] has no dependence on $\varepsilon$;
i.e., steric hindrances due to the adsorbed part of the polymer does
not cause the slowdown of adsorption at high
$\varepsilon$-values. Instead, as explained below quantitatively, the
physics of the slowdown of adsorption at larger $\varepsilon$-values
is explained by the lack of availability of not-yet-adsorbed monomers
near the surface.

For high adsorption energies the monomers that were close to the
surface at $t=0$ initially get quickly and effectively irreversibly
adsorbed, while the monomers that are far away from the surface at $t=0$
cannot respond to this fast change of the polymer's configuration near
the surface. As a result, during the adsorption process, the polymer
adopts a stretched configuration close to the surface, while far away
from the surface the polymer remains largely in its $t=0$ coiled
shape: this is the stem-flower picture of Brochard-Wyart
\cite{brochard95}. [It is precisely this stem-flower shape that
invalidates the physics behind the exponent $(1+\nu)/(1+2\nu)$,
seen at low adsorption energies. The number $(1+2\nu)$ in the
denominator is derived from the Rouse exponent, and the number
$(1+\nu)$ in the numerator assumes that during adsorption the
polymer's size scales as $\sim [N-n(t)]^\nu$; both fail at the stem
(of the stem-flower model), which is highly stretched.]
In fact, the occurrence of $\nu$ in the exponent $1/(1+\nu)$ at high
adhesion energies stems from the polymer's size-scaling $N^\nu$ at
$t=0$, as we argue next. Let us denote, by $z(t)$, the distance that
the stem extends in real space from the surface at time $t$. The total
number of monomers in the flower at time $t$ --- still largely in the
same coil shape as at $t=0$ --- is $Q(t)\sim N-z(t)^{1/\nu}$. In such
a configuration, the flower would lose monomers through the stem to
the surface, and the rate of loss of monomers is proportional to the
gradient of monomeric density along the stem, $\sim 1/z(t)$. The
solution of the differential equation $\dot Q(t)\sim 1/z(t)$ yields
$z(t)\sim t^{\nu/(1+\nu)}$. Since all the monomers that were present
within a distance $z(t)$ at $t=0$ --- apart from the few within the
stem at $z(t)$ --- are adsorbed by time $t$, the total number of
adsorbed monomers at time $t$ scales as $s(t)\sim z(t)^{1/\nu} \sim
t^{1/(1+\nu)}$. Note that in this qualitative derivation there is no
dependence on the adsorption energy [except that it needs to be
high!], as observed in the inset of Fig. \ref{fig2}(b). When
hydrodynamic interactions are included, following the physics of
field-driven translocation \cite{vocks08} we expect the adsorption
time to scale $\sim N^{(1+\nu)/(3\nu)}$ for not too high adhesion
energies; however, presently we do not understand how co-operative
motions of the monomers in the presence of hydrodynamics would affect
the exponent at high adhesion energies.

An increase in the energy of adsorption allows one to cross-over from
weak to strong regimes of physisorption, both characterised by
different exponents.  The remaining question is what determines the
energy scale $\varepsilon_{c}$ that separates these two regimes. This
can be understood if we return to $\Delta F \simeq - \varepsilon +
k_BT \ln{(\mu_{3}/\mu_{2})}$, wherein the first term lowers the free
energy due to favourable adsorption to the attractive surface, while
the second term increases the free energy because of entropy reduction
by going from three dimensions to a more constrained two-dimensional
surface. It is reasonable to suggest that two dynamic regimes of
adsorption are separated when the free energy gain per monomer is
comparable with thermal energy, i.e., $|\Delta F| \simeq k_BT$. Our
estimates for critical adsorption yield $\mu_{3}/\mu_{2}
\approx 3$, which leads to $\varepsilon_{c} \simeq 2$. Our simulations 
show that $\varepsilon_{c}\approx 5$, suggesting deviations from
(equilibrium) free-energy concepts in (non-equilibrium) surface
adhesion process. As shown in Fig. \ref{fig2}(c), the adsorption
proceeds faster for lower adsorption energies, with the most optimal
adsorption speed close to $\varepsilon_c$. Since these type of
energies are typical for protein-DNA interactions \cite{lodish_book},
one can suggest that this might be a mechanism by which biological
adhesion processes are controlled.

To conclude, using computer simulations and theoretical arguments we
studied single polymer adsorption to solid surfaces in the absence of
hydrodynamic interactions. Our analysis shows that the adhesion
process is non-equilibrium. Details of the adsorption process depend
on the strength of adsorption energies: for weak (polymer-surface)
interactions the dynamics is determined by memory effects as in
field-driven polymer translocation, while for strong interactions
adsorption is limited by stretching of the unadsorbed part of the
polymer. These two regimes are separated by the energy scale
that is determined by a balance between favourable enthalpic and
unfavourable entropic contributions due to adsorption of the monomers
to the surface. It is argued that the adsorption process is most
optimal at low interaction energies, and this might be the mechanism
by which biological surface adhesion processes are controlled.

D.P. gratefully acknowledges ample computer time on the Dutch national
super\-computer facility SARA. A.B.K. would like to acknowledge support
from the Welch Foundation (Grant C-1559), and the U.S. National Science
Foundation (Grant  NIRT ECCS-0708765).


\vspace{1cm}
\begin{thebibliography}{99}

\bibitem{fleer_book} C.J. Fleer {\it et al.}, {\it Polymers at
Interfaces} (Chapman and Hall, London 1993).

\bibitem{lodish_book} H. Lodish {\it et al.}, {\it Molecular Cell
Biology}, (W.H. Freeman and Company, New York, 2002), 4th Ed.

\bibitem{degennes87} P.G. de Gennes, Adv. Coll. Sci. {\bf 27}, 189
  (1987).

\bibitem{johnson92} H.E. Johnson and S. Granick, Science {\bf 255},
966 (1992); J.F. Douglas, {\it et. al.},  J. Phys. Cond. Matt. {\bf 9}, 7699
(1997); S. Minko, A. Voronov and E. Pefferkorn, Langmuir {\bf 16},
7878 (2000).

\bibitem{shaffer93} J.S. Shaffer and A.K. Chakraborty,  Macromolecules
{\bf 26}, 1120 (1993).

\bibitem{raviv02} U. Raviv, J. Klein and T.A. Witten, Eur. Phys. J. E
{\bf 9}, 405 (2002).

\bibitem{review05} B. O'Shaughnessy  and D. Vavylonis,
J. Phys. Cond. Matt. {\bf 17}, R63 (2005).

\bibitem{SV03} B. O'Shaughnessy  and D. Vavylonis, Eur. Phys. J. E {\bf 11}, 213 (2003).

\bibitem{shaffer94} J.S. Shaffer,  Macromolecules {\bf 27}, 2987
(1994).

\bibitem{ponomarev00} A.L. Ponomarev, T.D. Sewell, and C.J. Durning,
Macromolecules {\bf 33}, 2662 (2000).

\bibitem{oshaughnessy1} B. O'Shaughnessy  and D. Vavylonis,
Phys. Rev. Lett. {\bf 90}, 056103 (2003)

\bibitem{vaneijk98} M.C.P. Van Eijk, {\it et. al.},  Eur. Phys. J. B {\bf
1}, 233 (1998).

\bibitem{descas06} R. Descas, J. U. Sommer, and A. Blumen,
  J. Chem. Phys. {\bf 124}, 094701 (2006). 

\bibitem{bhatta08} S. Bhattacharya, {\it et al.}, Phys. Rev. E {\bf 77},
061603 (2008).

\bibitem{brochard95} F. Brochard-Wyart, Europhys. Lett. {\bf 30}, 387
  (1995).

\bibitem{vocks08} H. Vocks {\it et al.}, J. Phys. Cond. Matt. {\bf 20},
095224 (2008).

\bibitem{trans} D. Panja, G. T. Barkema and R. C. Ball, J. Phys.:
Condens. Mattter {\bf 19}, 432202 (2007); {\it ibid.} J. Phys.:
Condens. Matter {\bf 20}, 095224  (2008); D. Panja and G. T. Barkema,
Biophys. J. {\bf 94}, 1630 (2008).

\bibitem{model} A. van Heukelum and G. T. Barkema, J. Chem. Phys. {\bf
119}, 8197 (2003); A. van Heukelum {\it et al.}, Macromol. {\bf 36},
6662 (2003); J. Klein Wolterink {\it et al.}, Macromol. {\bf 38}, 2009
(2005).

\bibitem{klein_adsorbed} J. Klein Wolterink, G.T. Barkema and
  M.A. Cohen Stuart, Macromolecules {\bf 38}, 2009 (2005).

\bibitem{wolterink06} J.K. Wolterink, G.T. Barkema and D. Panja,
  Phys. Rev. Lett. {\bf 96}, 208301 (2006).

\bibitem{diehla} H. W. Diehla and B. Shpot, Nucl. Phys. B {\bf 528},
595 (1998).

\bibitem{luo08} A. Bhattacharya {\it et al.}, arXiv:0808.1868.

\bibitem{fyta} M. Fyta {\it et al.}, Phys. Rev. E {\bf 78}, 036704
  (2008). 

\end{thebibliography}
\end{document}